\documentclass[12pt]{article}

\usepackage[utf8]{inputenc}
\usepackage{indentfirst}
\usepackage{graphicx}
\usepackage{amsmath}
\usepackage{booktabs}
\usepackage{url}
\usepackage{authblk}
\usepackage[backend=biber,style=authoryear,natbib=true]{biblatex}
\usepackage{array}
\usepackage{enumitem}
\setlist{itemsep=0pt, parsep=0pt, topsep=2pt}
\usepackage[colorlinks=true,linkcolor=black,citecolor=black,urlcolor=black]{hyperref}
\title{A Faceted Proposal for Transparent Attribution of AI-Assisted Text Production  \\
\includegraphics[width=6cm]{myclass}}
\author{Geraldo Xexéo}
\affil{Programa de Engenharia de Sistemas e Computação\\
COPPE - Universidade Federal do Rio de Janeiro\\
Rio de Janeiro,RJ -- Brasil}
\date{}

\begin{document}

\maketitle

\begin{abstract}
Artificial intelligence systems are increasingly integrated into writing processes, challenging traditional notions of authorship, responsibility, and intellectual contribution.
Current disclosure practices usually indicate whether AI was used, but rarely explain how it was used, where it intervened, or how its output was reviewed.
This paper proposes a faceted model for representing AI-assisted text production at the levels of documents, chapters, sections, and paragraphs.
The proposal introduces a core model based on Form, Generation, and Evaluation, and an extended model that adds Intent, Control, and Traceability.
The model is positioned as a minimal operational baseline with extensibility toward higher-fidelity representations.
A worked example based on the production of this article demonstrates applicability.
\end{abstract}

\section{Introduction}

This paper is a position paper proposing a faceted model for the transparent attribution of AI-assisted text production.

Its central claim is that current declarations of generative AI use in academic and scientific writing are too coarse for the practical, ethical, and epistemic problems now faced by scholarly communication.

The paper does not argue that generative AI should be banned from writing workflows. Rather, it argues that AI involvement should be described with enough precision to distinguish correction, rewriting, generation, conceptual support, human control, evaluation, and traceability.

The intended contribution is therefore a structured vocabulary for disclosure, not a detector of AI-generated text and not a moral classification of acceptable or unacceptable use.

Generative AI has rapidly entered academic and scientific publication practices.Researchers now use large language models to draft paragraphs, revise manuscripts, summarize literature, translate text, create code, generate examples, prepare responses to reviewers, and explore conceptual framings.
This expansion has produced both opportunities and concerns.

On the opportunity side, AI systems can reduce linguistic barriers, accelerate routine drafting tasks, and help authors explore alternative formulations.

On the concern side, they can produce fabricated citations, inaccurate claims, misleading summaries, biased language, untraceable transformations, and ambiguity about intellectual contribution.

These problems have been discussed both in scientific literature and in public reporting. Early editorials in \textit{Science} and \textit{Nature} argued that AI systems should not be treated as authors because they cannot assume responsibility for scientific claims, approve final versions, or be accountable for errors \citep{thorp2023chatgpt, nature2023transparent}.

\citet{stokelwalker2023chatgpt} described both the enthusiasm and apprehension surrounding ChatGPT and related systems in scientific work, including the risk that generative systems may reshape research communication before norms and institutions are prepared.

A particularly concrete risk is bibliographic fabrication.
\citet{walters2023fabrication} showed that ChatGPT-generated scholarly texts can contain fabricated and erroneous citations, which is especially problematic because references are not merely decoration in scientific writing, but part of the evidence infrastructure of scholarship. As times passes, however, top level AI systems produce less errors. As of 2026, the bibliographic references are almost all correct with ChatGPT 5.4.

\citet{emsley2023fabrications} similarly argued that calling such outputs ``hallucinations'' may understate their significance in scientific contexts, where fabricated references and unsupported claims can function as falsifications of the scholarly record.

For the general public, journalistic reports have also documented the practical difficulty of detecting undisclosed AI use in academic journals and the appearance of fake or unverifiable references in scholarly materials \citep{wired2023academicai, times2025fakecitations}.

The institutional response has been fast but uneven. Publishers, societies, funders, and universities have generally converged on three principles: AI tools should not be authors, human authors remain responsible for final content, and relevant AI use should be disclosed. However, the level of detail required by these policies varies substantially.

 Publisher policies show a movement from prohibition toward conditional acceptance, but they still tend to rely on narrative declarations rather than structured, segment-level descriptions.
Elsevier allows AI-assisted technologies in manuscript preparation when used with human oversight and disclosure, but it emphasizes that authors must verify accuracy, check sources, preserve authentic intellectual contribution, and remain responsible for privacy, intellectual property, and rights issues \citep{elsevier2025genai}.
Springer Nature distinguishes AI-assisted copy editing from generative editorial work and requires human accountability for the final text, while also treating AI-generated images and videos cautiously because of unresolved copyright and research-integrity issues \citep{springernature2026ai}.

Scientific and professional societies have developed similar approaches. The Association for Computing Machinery states that generative AI tools may not be listed as authors, because authorship requires identifiable human responsibility, substantial intellectual contribution, and accountability for the published work. At the same time, ACM permits the use of generative AI tools to create content when that use is fully disclosed \citep{acm2025authorship}. IEEE guidance is even closer to the motivation of this paper because it requires that authors identify the AI system used, identify the specific sections that used AI-generated content, and briefly explain the level at which the AI system was used \citep{ieeeras2026genai}.

This is an important precedent, because it suggests that a useful disclosure should not merely say whether AI was used, but should indicate where and how it was used.

Ethics organizations have also influenced the emerging consensus.
COPE's position on authorship and AI tools states that AI tools cannot meet authorship requirements because they cannot take responsibility for submitted work \citep{cope2023aitools}.
This principle has been adopted or echoed by many publishers and societies.

However, saying that AI cannot be an author does not settle the harder question addressed by this paper: how should the contribution of AI be represented when it is neither authorship nor irrelevant tool use.

Universities and funders have also begun to formalize guidance.
Harvard's research guidance states that researchers should consult publisher rules, disclose generative AI use when required, and attend to risks such as bias, misinformation, intellectual-property compliance, security, unpredictability, and overreliance \citep{harvard2026genai}.
The Universidade Federal do Rio de Janeiro has published preliminary institutional documents on academic integrity and recommendations for AI use, emphasizing human intellectual responsibility, critical verification of AI outputs, transparency, intellectual property, and AI literacy \citep{ufrj2026ai}.

In Brazil, the Conselho Nacional de Desenvolvimento Científico e Tecnológico issued Portaria CNPq No. 2.664, of March 6, 2026, instituting a Policy of Integrity in Scientific Activity.
The purpose of that policy is to guarantee integrity in scientific activities supported by CNPq through education, prevention, investigation, and sanctions, while promoting ethics, transparency, responsible authorship, and integrity across the research lifecycle \citep{cnpq2026integridade}.
Its treatment of generative AI is especially significant for this paper. The policy does not prohibit generative AI use, but requires declaration of such use, regardless of the type of generative AI or the phase of research development, specifying the tool used and the purpose of use.
It also forbids submitting generative-AI-produced content as if it were human-authored and states that authors remain fully responsible for the final content, including plagiarism or inaccuracies generated by the tool.
In addition, it discourages the use of AI in the preparation of scientific reviews \citep{cnpq2026integridade}.

This policy is important because it moves beyond a purely publisher-centered view and frames AI use as part of scientific integrity, research governance, and institutional responsibility.
These responses are necessary, but still incomplete.

Most current policies require disclosure, but do not provide a sufficiently detailed language for disclosure.
They often do not distinguish whether AI was used for spelling correction, stylistic revision, translation, summarization, generation of new content, conceptual framing, bibliographic formatting, or evaluation.
They rarely specify the scale of use, such as document, chapter, section, subsection, paragraph, figure, table, code fragment, or reference list.
They also tend to under-specify traceability, including whether prompts, logs, model versions, intermediate drafts, or human review records are available.
This creates a practical gap between the demand for transparency and the means available to express transparency.

The premise of this paper is that generative AI should be understood as a new cognitive tool.
In this respect, it belongs to a long history of instruments that extend human capacities. Lenses extended perception. calculators extended arithmetic reasoning, computers and internet extended symbolic manipulation, storage, search, simulation and communication.
Now, generative AI extends linguistic production, conceptual exploration, reformulation, summarization, translation, and argumentative drafting.

The appropriate response is, therefore, not nostalgia for a pre-AI writing environment, but the construction of norms that make AI-assisted cognition visible, accountable, and interpretable.

However, generative AI differs conceptually from many earlier tools. It operates in domains historically associated with human cognition, such as writing, reasoning, explanation, design, and artistic creation. These activities have been treated not merely as outputs, but as signs of understanding, authorship, originality, and intellectual identity.

This difference explains why AI use in writing raises stronger concerns than the use of a calculator or a spelling checker.
The issue is not only efficiency, but the distribution of intellectual contribution. 

The first concern seems to be intellectual property. The U.S. Copyright Office has argued that copyright protection for AI-assisted outputs depends on sufficient human authorship and expressive control, and that prompts alone may be insufficient when they do not determine expressive elements of the output \citep{usco2025copyrightability}. WIPO has similarly emphasized that generative AI challenges traditional ideas of authorship, originality, training-data rights, and ownership of generated outputs \citep{wipo2024genaiip}.

A disclosure model should therefore help distinguish cases in which AI merely assists expression from cases in which it contributes substantially to expressive or conceptual form.

The second concern is research ethics. Ethical scholarly communication depends on responsibility, attribution, honesty, and the possibility of correction. If AI-generated content is incorporated without disclosure, readers, reviewers, and institutions may misinterpret the origin of claims, the labor of authors, and the reliability of the work.
This is why policies from COPE, ACM, Elsevier, IEEE, CNPq, and universities converge on human responsibility and disclosure \citep{cope2023aitools, acm2025authorship, elsevier2025genai, ieeeras2026genai, cnpq2026integridade}.

The third concern is transparency. A statement such as ``AI was used in the preparation of this manuscript'' is often true but weak. It does not say whether AI shaped the argument, generated examples, rewrote sections, translated text, formatted references, or checked grammar. It also does not say whether the author reviewed the output, preserved prompts, or verified sources.
A structured model can make disclosure more informative without requiring every text to include a full production log.

The fourth concern is bias. Large language models may reproduce or amplify biases present in training data, system design, reinforcement processes, or user prompts. 
This problem has been discussed in the broader literature on language-model risks, including concerns about representational harms, stereotyping, exclusion, and the appearance of neutrality in outputs that are in fact socially and historically patterned \citep{bender2021dangers, weidinger2021ethical}.
In academic writing, bias may affect literature summaries, terminology, examples, framing of controversies, and treatment of marginalized groups.

The fifth concern is error propagation. Generative AI can produce fluent text that contains false statements, fabricated sources, incorrect attributions, or unsupported generalizations. 
Because such outputs often appear coherent, errors can propagate from drafts into manuscripts, from manuscripts into publications, and from publications into future literature reviews.
The problem is especially serious when AI-generated references, claims, or summaries are accepted without verification \citep{walters2023fabrication, emsley2023fabrications, sun2024hallucination}.

These concerns motivate the model proposed in this paper.

The model is faceted because no single scale can capture the relevant distinctions. A text segment may be AI-edited but human-authored. It may be AI-generated but fully reviewed. It may be generated through a long conversation but remain strongly human-controlled. It may be traceable but poorly evaluated.
It may involve conceptual support without containing any sentence directly copied from an AI output.

The following sections therefore introduce a core model based on Form, Generation, and Evaluation, followed by an extended model that adds Intent, Control, and Traceability.

The paper then discusses application across textual scales, identifies AI systems as referenceable tools, and presents a worked example based on the production of this article.

\section{Related Work}

The problem addressed in this paper is situated at the intersection of provenance, authorship, publication ethics, research integrity, educational assessment, and content authenticity.
Recent work does not converge on a single model.
Instead, different communities have proposed different kinds of solutions, each motivated by a different practical problem.

This section groups related work according to the generic intent of each group.
The comparison shows that existing approaches cover parts of the problem, but none provides a compact, multi-scale, text-segment-oriented model covering Form, Generation, Evaluation, Intent, Control, and Traceability.

\subsection{Provenance and Contribution Models}

The first group consists of general provenance and contribution models.
Their generic intent is to represent who or what contributed to an artifact, what activities were involved, and how responsibility should be attributed.

The W3C PROV model is the most important generic provenance framework.
It represents provenance through entities, activities, and agents \citep{moreau2013prov}.
This is directly relevant because AI-assisted text production can be interpreted as a sequence of activities involving human authors, AI systems, prompts, drafts, revisions, and outputs.
However, PROV is deliberately domain-general.
It does not provide categories for distinguishing spelling correction, grammatical revision, conversational generation, conceptual support, or human review.

The CRediT taxonomy addresses a different but related problem.
It standardizes human contributor roles, including conceptualization, methodology, writing original draft, and writing review and editing \citep{allen2014credit, brand2015beyond}.
CRediT is useful because it separates kinds of intellectual contribution.
However, it was designed for human contributors, not for AI systems.
It also classifies roles at the contribution level rather than transformations at the level of documents, sections, or paragraphs.

Model Cards and Datasheets belong to the same broad family of documentation frameworks.
Model Cards document model behavior, intended use, limitations, and evaluation conditions \citep{mitchell2019model}.
Datasheets for Datasets document datasets, their motivation, composition, collection process, preprocessing, uses, and limitations \citep{gebru2021datasheets}.
These frameworks are important because they show that responsible use of AI requires structured contextual documentation.
However, they document systems and datasets rather than the role of AI in a specific text segment.

These models motivate the present proposal but do not replace it.
PROV explains how to represent provenance in general.
CRediT explains how to represent human contribution.
Model Cards and Datasheets explain how to document AI-related resources.
The present model focuses on a narrower question: how to describe the role of AI in the production of specific textual units.

\subsection{Publisher, Journal, and Society Policies}

The second group consists of publisher, journal, and society policies.
Their generic intent is normative and procedural.
They define what authors may do, what must be disclosed, and who remains responsible for the final work.

COPE states that AI tools cannot be credited as authors because they cannot take responsibility for submitted work \citep{cope2023aitools}.
Elsevier allows generative AI and AI-assisted technologies in manuscript preparation when authors provide oversight, verify outputs, and disclose use, while maintaining human responsibility for the final content \citep{elsevier2025genai}.
PLOS requires authors to disclose AI tools used in the study or article content and to ensure the accuracy and correct attribution of ideas generated by AI tools \citep{plos2023ai}.
ACM similarly treats authorship as a human responsibility and requires disclosure of generative AI use when it contributes to submitted material \citep{acm2025authorship}.
IEEE-related guidance requires authors to identify the AI system used, identify the affected sections, and describe the level of use \citep{ieeeras2026genai}.

These policies are necessary because they establish accountability.
They converge on three principles.
AI systems are not authors.
Human authors remain responsible.
Relevant AI use must be disclosed.

However, policies usually do not provide a sufficiently expressive classification of AI involvement.
A statement such as ``AI was used for manuscript preparation'' does not distinguish language correction from conceptual framing.
Nor does it distinguish a single prompt from a long multi-turn process.
Nor does it specify whether prompts, logs, model identifiers, or intermediate drafts are available.

For this reason, publication policies motivate the need for a structured annotation model.
They create the obligation to disclose, but usually do not provide a precise vocabulary for what should be disclosed.

\subsection{Structured AI Disclosure Frameworks}

The third group consists of structured AI disclosure frameworks.
Their generic intent is to make disclosure more systematic, comparable, and easier to produce.

The Artificial Intelligence Disclosure Framework, or AID, is one of the closest proposals to the present work.
It provides a structured statement for reporting AI use across research and writing activities \citep{weaver2024aid, weaver2026aid}.
AID includes categories such as tool identification, conceptualization, methodology, information collection, data analysis, interpretation, visualization, writing review and editing, and translation.
Its goal is to make AI-use reporting clear, consistent, succinct, and partially machine-readable.

DAISY is a recent evaluated disclosure tool for academic writing \citep{daisy2026}.
It operationalizes disclosure as a form-based author-facing interaction.
It was developed from literature-derived requirements and co-design, and then evaluated with authors.
DAISY is especially important because it shows that structured disclosure can improve completeness without necessarily reducing author comfort.
It also frames AI disclosure as a sociotechnical practice rather than merely as a compliance statement.

GAIDeT is another recent proposal aimed at structured disclosure of generative AI use in research and publishing \citep{gaidet2025}.
Its intent is pragmatic: reduce vagueness in declarations such as ``ChatGPT was used to improve clarity'' and give researchers a more precise way to state what AI contributed.
Although it is less formal than PROV and less empirically evaluated than DAISY, it reflects the same movement toward structured, activity-aware disclosure.

Resnik and Hosseini argue that AI-use disclosure should not be indiscriminate.
They distinguish mandatory, optional, and unnecessary disclosure and argue that disclosure should depend on whether AI use is intentional, significant, and relevant to evaluating the work \citep{resnik2025disclosing}.
This position is important because it cautions against excessive disclosure burden.
It also supports the distinction made in this paper between a minimal core model and a more detailed extended model.

Compared with these frameworks, the present proposal differs in three ways.
First, it is explicitly faceted.
Second, it is designed for multiple textual scales, from document to paragraph.
Third, it distinguishes Form, Generation, Evaluation, Intent, Control, and Traceability as separate dimensions, rather than treating AI use as a single disclosure narrative.

\subsection{Manuscript-Preparation Classification Proposals}

The fourth group consists of classification proposals focused specifically on manuscript preparation.
Their generic intent is to help publishers, editors, reviewers, and authors distinguish acceptable, reportable, and problematic forms of AI assistance.

The STM Association published ethical and practical guidelines for the use of generative AI in scholarly communications in 2023 \citep{stm2023genai}.
In 2025, STM released a draft report on classifying AI use in manuscript preparation, developed by its Task and Finish Group on AI Labelling Terminology \citep{stm2025classification}.
This draft explicitly responds to the fact that publisher guidelines have not kept pace with changing AI practices.
It aims to give publishers a clearer framework for defining, evaluating, and guiding transparent AI use in manuscript preparation.

This family of proposals is close to the present work because it is concerned with the manuscript as an object of production.
However, such classifications tend to be policy-facing.
They are intended to help publishers define rules.
The model proposed here is more text-facing.
It is intended to annotate documents, sections, subsections, and paragraphs in a way that can be read by humans and, eventually, processed by tools.

\subsection{Educational AI-Use Scales}

The fifth group consists of educational frameworks for declaring or regulating AI use in assessment.
Their generic intent is not publication transparency, but academic integrity and assessment design.

The Artificial Intelligence Assessment Scale, or AIAS, is the most influential example \citep{perkins2024aias}.
It defines levels of permitted AI use in educational assessment, from no AI assistance to full AI integration.
Its purpose is to help educators communicate expectations and align AI use with learning outcomes.
The updated versions of AIAS emphasize that AI can be integrated ethically when expectations are explicit and when assessment design reflects the intended learning process.

AIAS is relevant because it treats AI use as a matter of levels, not as a binary condition.
It also shows that disclosure and permission must be contextual.
However, AIAS is designed for assessment settings, not for scholarly publication.
It focuses on what students are allowed to do, not on how a published text should represent the history of its production.

The present model borrows the idea that AI use can be level-based.
However, it replaces a single educational scale with multiple facets that better fit research writing and publication.

\subsection{Content Authenticity, Metadata, and Detection Approaches}

The sixth group consists of content authenticity and detection approaches.
Their generic intent is to establish whether content has been generated, edited, or authenticated, usually through metadata, signatures, watermarking, or forensic detection.

The C2PA specification provides a technical framework for content provenance and authenticity metadata \citep{c2pa2024}.
Its purpose is to support verifiable claims about the origin and transformation of digital content.
In visual and multimedia domains, this approach is especially important because metadata and signed manifests can help users assess whether content has been created or modified.

Detection-oriented work takes a different path.
It attempts to infer whether text or reviews were AI-modified.
For example, recent work has monitored AI-modified content at scale in peer-review contexts \citep{liang2024monitoring}.
Journalistic reports and empirical studies also suggest that AI use in academic writing is difficult to detect reliably and that undisclosed use remains a practical challenge \citep{knibbs2023academicai, perlis2026disclosure}.

These approaches are important, but they solve a different problem.
Detection asks whether AI may have been used.
Content authenticity asks whether provenance metadata can be verified.
The present proposal asks how authors can declare the role of AI in a meaningful and structured way.
For this reason, the model is complementary to C2PA and detection tools.
It could eventually be embedded in metadata systems, but its primary contribution is conceptual and documentary.

\subsection{Comparison of Related Approaches}

Table~\ref{tab:related-comparison} summarizes how the main groups of related work address the six facets proposed in this paper.

\begin{table}[!hbt]
\centering
\footnotesize
\begin{tabular}{>{\raggedright\arraybackslash}p{2cm}>{\raggedright\arraybackslash}p{2cm}cccccc>{\raggedright\arraybackslash}p{3cm}}
\toprule
Group & Examples & F & G & E & I & C & T & Additional emphasis \\
\midrule
Provenance and contribution models
& PROV, CRediT, Model Cards, Datasheets
& P & P & P & P & P & P
& General provenance, contributor roles, model and dataset documentation \\

Publisher and society policies
& COPE, Elsevier, PLOS, ACM, IEEE
& P & P & P & P & P & P
& Authorship responsibility, disclosure obligation, editorial compliance \\

Structured disclosure frameworks
& AID, DAISY, GAIDeT
& P & P & P & S & P & P
& Disclosure statements, activity categories, author-facing forms \\

Manuscript-preparation classifications
& STM 2023 guidance, STM 2025 draft classification
& P & S & P & P & P & P
& Publisher-facing terminology for AI use in manuscript preparation \\

Educational AI-use scales
& AIAS
& P & S & P & S & P & N
& Permitted use in assessment, academic integrity, learning outcomes \\

Authenticity and detection approaches
& C2PA, watermarking, AI-use detection
& P & P & N & N & N & S
& Verifiable metadata, content authenticity, forensic detection \\

Present proposal
& Core and extended faceted model
& S & S & S & S & S & S
& Multi-scale text annotation, document to paragraph level \\
\bottomrule
\end{tabular}
\caption{Comparison of related approaches against the six facets proposed in this paper.
S indicates strong explicit coverage.
P indicates partial or indirect coverage.
N indicates that the facet is not a central concern.}
\label{tab:related-comparison}
\end{table}

The comparison shows that existing work is complementary rather than redundant.
Provenance frameworks offer general structure, but not writing-specific levels.
Publisher policies create disclosure obligations, but not detailed annotation schemes.
Structured disclosure frameworks are close to the present proposal, but usually focus on disclosure statements rather than segment-level annotation.
Educational scales provide useful precedents for level-based AI use, but are oriented toward assessment rather than publication.
Authenticity frameworks support verification, but do not explain the semantic role of AI in writing.

The gap addressed by this paper is therefore specific.
It proposes a compact, faceted, multi-scale model for describing how AI participates in the production of text.
\section{Core Facets}

The core model contains the minimum set of facets needed to describe AI-assisted text production in a practical way.
It answers three basic questions.

First, what happened to the wording and structure of the text.
Second, how the text was produced.
Third, how the resulting text was reviewed.

These questions correspond to three facets: Form, Generation, and Evaluation.
They are deliberately separated because they describe different properties of a text.
A paragraph may be human-authored but AI-edited.
Another paragraph may be AI-generated but fully reviewed by a human expert.
A third paragraph may be AI-generated and left unrevised.
A single label such as ``AI-generated'' would fail to distinguish these cases.

\subsection{Form}

The Form facet describes transformations applied to a text after an initial version already exists.
It concerns spelling, punctuation, grammar, style, fluency, terminology, organization, and rhetorical structure.

This facet is necessary because different forms of editing have different consequences for meaning and authorship.
Correcting a spelling error does not normally change the intellectual content of a text.
Reorganizing the argument of a section may change how the reader understands the contribution.
For this reason, the Form facet is defined as a scale of increasing textual transformation.

This facet is related to editorial work and to the ``Writing -- review and editing'' role in the CRediT taxonomy \citep{allen2014credit, brand2015beyond}.
However, CRediT identifies contributor roles, while the present model classifies the intensity and scope of transformation.
This distinction is important because the same contributor role can include very different kinds of intervention.

The proposed levels are:

\begin{itemize}
\item F0: no form modification.
\item F1: orthographic correction.
\item F2: grammatical correction.
\item F3: local stylistic refinement.
\item F4: global rhetorical or structural restructuring.
\end{itemize}

F1 covers spelling, punctuation, capitalization, and typographic normalization.
F2 covers grammatical repair at the sentence level.
F3 covers local improvements in clarity, fluency, concision, terminology, or tone, without changing the overall argumentative structure.
F4 covers reordering paragraphs, restructuring sections, changing the explanatory sequence, merging or splitting arguments, or altering the rhetorical architecture of the text.

The boundary between F3 and F4 is especially important.
F3 improves the expression of a local idea.
F4 changes the arrangement of ideas.
Thus, F4 has stronger implications for interpretation and intellectual contribution.

\subsection{Generation}

The Generation facet describes how a text segment came into existence.
It asks whether the segment was written by a human, completed with AI assistance, generated from a prompt, developed through conversation, or produced by a structured pipeline.

This facet is related to the concept of activities in the W3C PROV model \citep{moreau2013prov}.
In PROV, an entity is produced by an activity involving agents and inputs.
However, PROV does not distinguish the specific interactional forms of AI-assisted writing.
The Generation facet refines that space for text production.

The proposed levels are:

\begin{itemize}
\item G0: fully human-authored.
\item G1: AI-assisted completion.
\item G2: AI-generated with a simple prompt.
\item G3: AI-generated with a detailed prompt.
\item G4: AI-generated through iterative conversation.
\item G5: AI-generated through a structured multi-step conversational or computational pipeline.
\end{itemize}

G0 means that the segment was produced without generative AI.
G1 describes autocomplete, phrase completion, or small local suggestions.
G2 describes generation from a short or underspecified prompt.
G3 describes generation from a prompt that specifies audience, tone, structure, constraints, examples, references, or argumentative goals.
G4 describes a multi-turn process in which the author questions, corrects, redirects, expands, or rejects model outputs.
G5 describes a designed workflow, such as multiple prompts, retrieval-augmented generation, multi-agent interaction, tool use, staged rewriting, or scripted generation.

Conversationality is explicitly represented because many AI-written texts are not produced by one prompt.
They emerge through negotiation.
The author may test alternatives, ask for criticism, demand restructuring, and progressively constrain the output.
This makes G4 and G5 qualitatively different from simple prompt generation.

\subsection{Evaluation}

The Evaluation facet describes how the resulting text was reviewed, checked, or validated.
It is separate from Generation because the reliability of a text is not determined only by how it was produced.
An AI-generated section may be carefully reviewed.
A human-authored section may contain unchecked errors.

This facet is motivated by concerns in model documentation, quality assurance, and scientific review.
Model Cards emphasize that AI systems should be documented with respect to performance, limitations, and evaluation conditions \citep{mitchell2019model}.
The Evaluation facet applies a similar concern to textual outputs.

The proposed levels are:

\begin{itemize}
\item E0: no revision.
\item E1: automated review only.
\item E2: partial human review.
\item E3: full human review.
\item E4: multi-stage or independent validation.
\end{itemize}

E0 means that the text was accepted without review.
E1 means that only automated tools were used, such as grammar checkers, spellcheckers, plagiarism detectors, or AI evaluators.
E2 means that a human reviewed selected parts or selected aspects of the text.
E3 means that a human reviewed the entire segment.
E4 means that the segment underwent structured validation, such as review by multiple people, review by a domain expert, independent checking of claims, or documented editorial stages.

Evaluation must remain distinct from both Form and Generation.
Form describes transformation.
Generation describes production.
Evaluation describes validation.
Keeping these facets separate prevents a common confusion: a polished text is not necessarily a reviewed text, and a reviewed text is not necessarily human-authored.

\subsection{Core Representation}

The core model represents a text segment as:

\begin{equation}
S = (F, G, E)
\end{equation}

This representation is intentionally compact.
It is suitable for lightweight disclosure in articles, dissertations, reports, teaching materials, and institutional documents.
It can be applied at the level of a document, chapter, section, subsection, or paragraph.

The core model is therefore a minimal operational baseline.
It is not intended to capture every relevant issue.
Rather, it provides a stable foundation on which additional facets can be added.

\section{Extended Facets}

The extended model adds facets that are not strictly necessary for minimal disclosure but are important for high-fidelity annotation.
These facets address questions that the core model leaves under-specified.

Why was AI used.
Who controlled the process.
Can the process be reconstructed.

These questions correspond to Intent, Control, and Traceability.
They are introduced as extensions because they increase expressive power but also increase annotation burden.

\subsection{Intent}

The Intent facet describes why AI was used.
This facet is necessary because the same technical interaction may serve very different scholarly functions.
A prompt may be used to correct grammar, translate a paragraph, summarize a source, generate a draft, or help design a conceptual framework.

Intent is related to CRediT because CRediT distinguishes roles such as conceptualization, writing, reviewing, supervision, and methodology \citep{allen2014credit, brand2015beyond}.
It is also related to Datasheets for Datasets, which emphasize the importance of documenting intended use \citep{gebru2021datasheets}.
In the present model, however, intent is applied to text segments rather than to datasets or human contributors.

The proposed levels are:

\begin{itemize}
\item I0: no AI intent or unspecified use.
\item I1: linguistic correction.
\item I2: transformation of existing text.
\item I3: generation of new textual content.
\item I4: conceptual support, ideation, framing, or argument development.
\end{itemize}

I1 covers spelling, grammar, and style assistance.
I2 covers translation, summarization, paraphrasing, adaptation, or rewriting of existing text.
I3 covers the production of new prose, examples, explanations, or drafts.
I4 covers conceptual work, such as proposing categories, structuring an argument, identifying gaps, generating research questions, or helping define a model.

I4 is especially important because it touches the traditional boundary of intellectual contribution.
Even if a human rewrites every sentence, AI may still have influenced the conceptual architecture of the work.
For this reason, Intent should not be collapsed into Generation.
Generation tells us how text was produced.
Intent tells us what role AI played in the author's intellectual workflow.

\subsection{Control}

The Control facet describes who directed the production process.
It is not redundant with Generation.
Generation describes the mechanism by which text was produced.
Control describes agency, constraint, and decision authority.

This distinction is motivated by the separation between agents and activities in PROV \citep{moreau2013prov}.
An activity may generate a text, but attribution also depends on who selected the goals, imposed constraints, accepted outputs, and made final decisions.

The proposed levels are:

\begin{itemize}
\item C0: fully human-controlled.
\item C1: AI used under narrow constraints.
\item C2: guided human-AI interaction.
\item C3: AI-dominant production with human selection or approval.
\item C4: largely autonomous pipeline.
\end{itemize}

C0 means that AI did not direct the content.
C1 means that AI was used for narrowly bounded operations.
C2 means that the human author actively guided the interaction through constraints, corrections, and selection.
C3 means that the AI produced most of the relevant content, while the human mainly selected, accepted, or lightly modified the result.
C4 means that the process was largely automated, such as a pipeline that generates, revises, evaluates, and formats text with limited human intervention.

Control is essential for distinguishing assistance from delegation.
Two texts may both be produced through conversation, but in one case the author may strongly guide every step, while in another the model may effectively determine the structure and content.
This difference matters for authorship, accountability, and evaluation.

\subsection{Traceability}

The Traceability facet describes whether the production process can be reconstructed.
It concerns prompts, logs, model identifiers, intermediate versions, metadata, and external verification.

This facet is related to C2PA, which addresses content provenance and authenticity through metadata and verification mechanisms \citep{c2pa2024}.
It is also connected to reproducibility in scientific work.
A claim about AI use is stronger when the process can be inspected, repeated, or audited.

The proposed levels are:

\begin{itemize}
\item T0: no traceability.
\item T1: partial informal notes.
\item T2: prompts or excerpts available.
\item T3: full logs, model identification, and intermediate versions available.
\item T4: independently verifiable provenance.
\end{itemize}

T0 means that no record of the process exists.
T1 means that the author kept informal notes, but not enough to reconstruct the process.
T2 means that prompts, relevant excerpts, or partial interaction records are available.
T3 means that the workflow includes full logs, model names or versions, dates, intermediate drafts, and relevant settings.
T4 means that provenance is independently verifiable, for example through signed metadata, repository history, authenticated logs, or trusted third-party records.

Traceability should not be confused with truthfulness.
A traceable process can still produce poor text.
However, traceability makes it possible to evaluate, audit, and contest the process.

\subsection{Extended Representation}

The extended model represents a text segment as:

\begin{equation}
S = (F, G, E, I, C, T_r)
\end{equation}

The subscript in $T_r$ distinguishes Traceability from the text segment itself.

The extended representation is suitable when the annotation must support auditability, research, education, publication policy, or high-stakes accountability.
It is more demanding than the core model, but it captures distinctions that are important when AI use affects conceptual contribution, reproducibility, or institutional responsibility.

The extended model should not be treated as mandatory in all contexts.
A short classroom essay may only require the core model.
A journal article, dissertation, technical report, or policy document may require selected extended facets.
A fully auditable workflow may require all extended facets.

\subsection{Discussion of Extension Costs}

The main advantage of the extended model is fidelity.
It can distinguish correction from conceptual support, generation from control, and review from traceability.
These distinctions are important because they correspond to different risks and responsibilities.

The main disadvantage is annotation cost.
More facets require more decisions.
More decisions can produce inconsistency, fatigue, and false precision.
If authors cannot reliably distinguish adjacent levels, the model may appear more rigorous than it actually is.

For this reason, the proposal separates the core model from the extended model.
The core model provides adoption.
The extended model provides depth.
A community standard can then define which facets are required, recommended, or optional for different document types.

\subsection{Redundancy Analysis}

The extended facets may appear redundant at first, but they capture different questions.

Generation and Control are distinct.
Generation asks how the text was produced.
Control asks who directed the production.

Evaluation and Traceability are distinct.
Evaluation asks whether the text was reviewed.
Traceability asks whether the production process can be reconstructed.

Intent and Generation are distinct.
Intent asks why AI was used.
Generation asks by what process the segment was produced.

These distinctions prevent conceptual collapse.
A text can be generated through conversation but remain strongly human-controlled.
A text can be generated from a simple prompt but used for conceptual ideation.
A text can be fully reviewed but have no preserved logs.
A text can have complete logs and still contain factual errors.

The extended model is therefore not merely a longer version of the core model.
It captures additional dimensions that become relevant when the goal is not only disclosure but also accountability, reproducibility, and interpretation.

\section{Representations of the Annotation}

The proposed model separates the semantic annotation from its visual or textual rendering. This separation is important because the same annotation may be used in different contexts:
\begin{itemize}
    \item A paper submitted to a publisher may need a prose disclosure statement. 
    \item A teaching document may need visible marks next to paragraphs. 
    \item A repository or metadata system may need a compact, machine-readable representation.

    \item A printed article may use icons when space allows, but a text-only version must remain available for accessibility, indexing, and institutional compliance.

\end{itemize}

For this reason, the model supports three complementary forms of representation: formal representation, iconic representation, and textual representation.

\subsection{Formal Representation}

The formal representation defines the annotation as a tuple assigned to a textual segment.
A segment may be a document, chapter, section, subsection, paragraph, table, figure, code fragment, or reference list.

Let $s$ be a textual segment.
The core annotation is defined as:

\begin{equation}
A_c(s) = (F, G, E)
\end{equation}

where $F$ is the Form level, $G$ is the Generation level, and $E$ is the Evaluation level.

The extended annotation is defined as:

\begin{equation}
A_e(s) = (F, G, E, I, C, T_r)
\end{equation}

where $I$ is the Intent level, $C$ is the Control level, and $T_r$ is the Traceability level.

The subscript in $T_r$ avoids ambiguity with $T$ as a possible symbol for text or time.

A complete annotation may also include scope and source information.
\begin{equation}
A(s) = (scope, F, G, E, I, C, T_r, tools, evidence)
\end{equation}

The field $scope$ identifies the level of the document structure to which the annotation applies.
The field $tools$ identifies the AI systems used.
The field $evidence$ identifies the level or location of supporting material, such as prompts, logs, drafts, metadata, or disclosure notes.

For example, a document-level annotation for the present article may be written as:
\begin{equation}
A(document) = (document, F4, G4, E3, I4, C2, T2)
\end{equation}

This means that the document underwent global restructuring, was generated through iterative conversation, received full human review, involved conceptual support, was guided by human-AI interaction, and has partial traceability through the available conversation record.

The formal representation is useful because it is compact, compositional, and suitable for computational processing.
It can be attached to document metadata, represented in JSON, encoded in LaTeX commands, or exported to provenance systems.

\subsection{Iconic Representation}

The iconic representation renders each facet-level pair as a visual symbol. Its purpose is rapid recognition.
Instead of requiring the reader to parse a full disclosure paragraph at every occurrence, the document can display a compact sequence of icons.

For example, a section may be marked with icons corresponding to:
\begin{equation}
(F4, G4, E3, I4, C2, T3)
\end{equation}

The iconic representation is appropriate when visual communication is useful, such as in textbooks, reports, preprints, teaching materials, institutional documents, or authoring environments. It is also useful for paragraph-level annotation, where repeated full prose disclosures would be intrusive.

\begin{figure}
    \centering
    \includegraphics[width=0.8\linewidth]{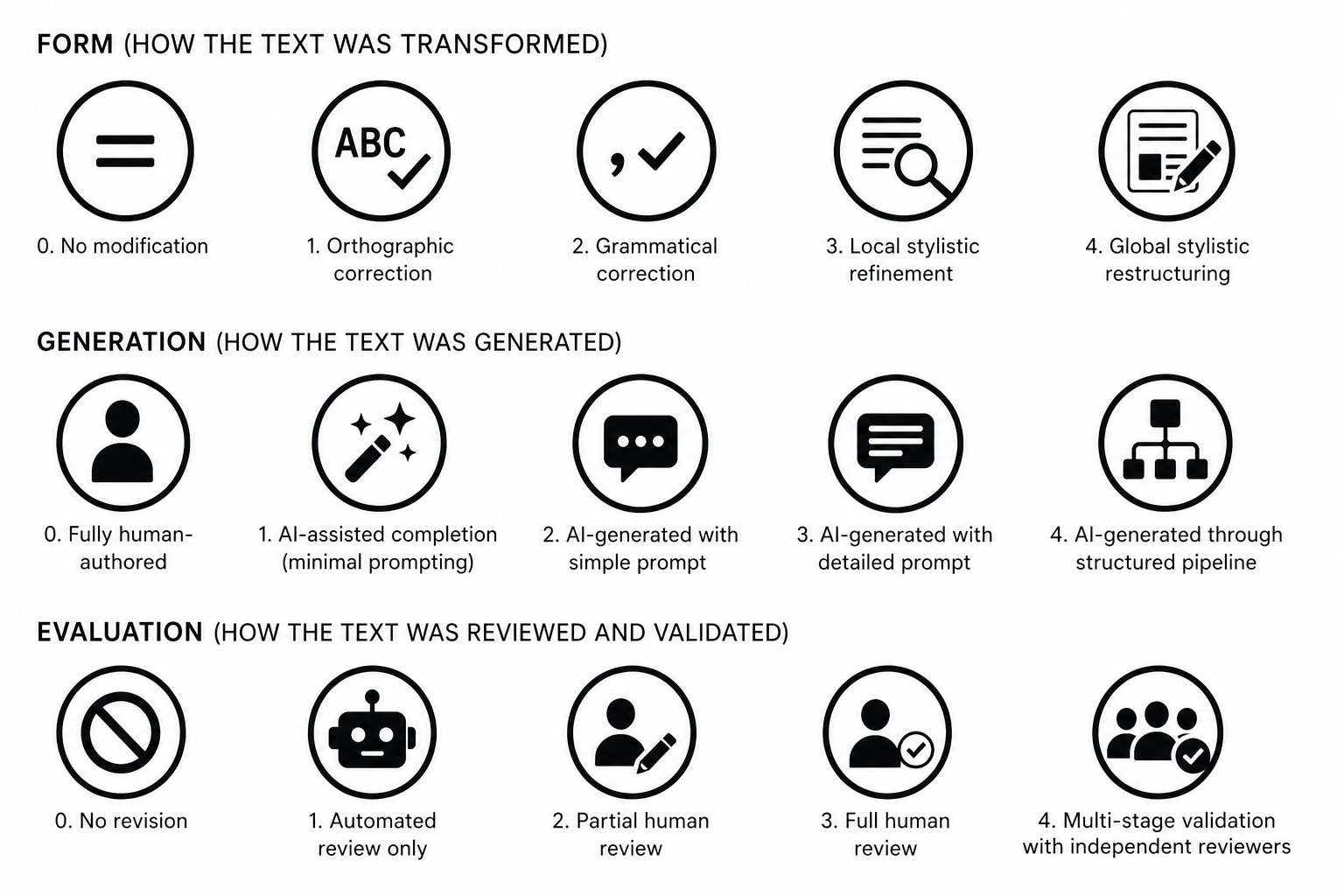}
    \caption{Proposed icons for the core model.}
    \label{fig:core}
\end{figure}

\begin{figure}
    \centering
    \includegraphics[width=0.8\linewidth]{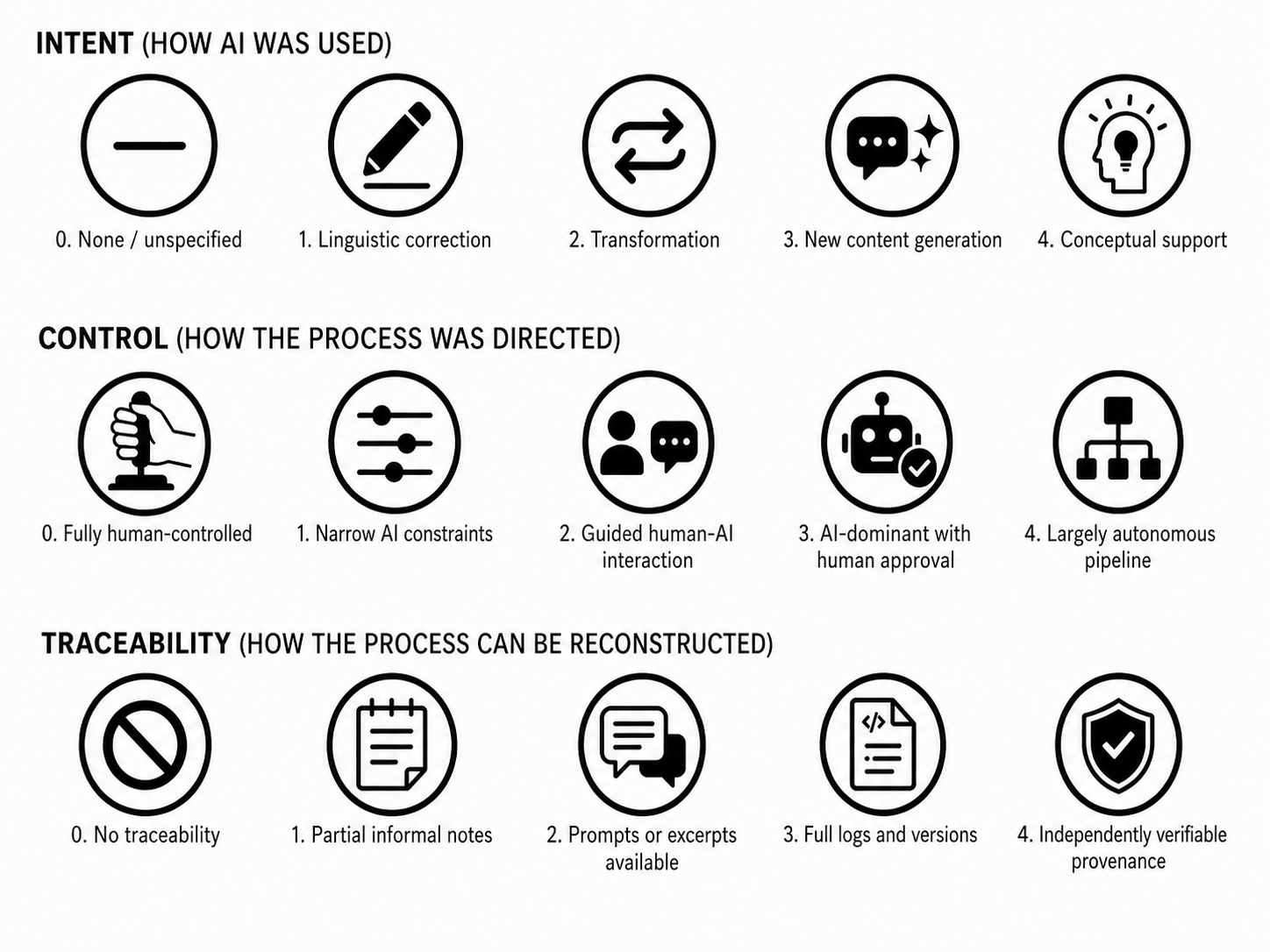}
    \caption{Proposed icons for the extended model}
    \label{fig:ext}
\end{figure}

\autoref{fig:core} shows a proposal for icons for the core model, while \autoref{fig:ext} shows a proposal for the extensions.

However, icons should not be the only representation.
Icons depend on reader familiarity, may not be accessible to screen readers, may be lost in plain-text exports, and may be ambiguous without a legend.

For this reason, every iconic mark should have a textual equivalent. The proposed iconic system should therefore be treated as a rendering layer, not as the primary semantic model. The semantic model is the tuple. The icon is one possible display of that tuple.

\subsection{Textual Grid Representation}

The textual grid representation is a compact text-only substitute for the iconic representation.
It is designed for accessibility, plain-text environments, metadata fields, institutional forms, and documents where icons are not desired or not supported.

A fully extended annotation can be represented as:

\begin{center}
\begin{tabular}{cccccc}
\toprule
Form & Generation & Evaluation & Intent & Control & Traceability \\
\midrule
F4 & G4 & E2 & I4 & C2 & T2 \\
\bottomrule
\end{tabular}
\end{center}

The same information can be written inline as:

\begin{center}
\texttt{|F4|G4|E2|I4|C2|T2|}
\end{center}

This notation is deliberately simple; it can be read by humans, copied into metadata, and used in LaTeX comments, Markdown documents, HTML attributes, institutional forms, or repository descriptions. It can also substitute for the iconic form when the icons are unavailable.

If future extensions require qualifiers, suffixes may be added after the level. For example, \texttt{I4Z} could indicate an institution-specific variant of conceptual support.

However, the baseline model should avoid suffixes unless they are defined in a local extension or a future standard.

The standard representation should, therefore, prefer:
\begin{center}
\texttt{|F4|G4|E2|I4|C2|T2|}
\end{center}
rather than ambiguous local codes.

A grid may also be attached to a specific scope:
\begin{center}
\texttt{section: |F3|G4|E2|I3|C2|T2|}
\end{center}
or:
\begin{center}
\texttt{paragraph 12: |F2|G1|E3|I1|C1|T1|}
\end{center}

This makes the representation usable at the document, section, subsection, and paragraph levels.

\subsection{Automatically Generated Textual Disclosure}

The textual grid can be expanded into a prose disclosure statement.

This is useful for publishers, funding agencies, universities, journals, and conference management systems that require human-readable declarations.

For example, the annotation:

\begin{center}
\texttt{|F4|G4|E2|I4|C2|T2|}
\end{center}

can be expanded into:

\begin{quote}
This text underwent global rhetorical or structural restructuring.
It was produced through iterative human-AI conversation.
It received partial human review.
AI was used for conceptual support, ideation, framing, or argument development.
The process was guided by human-AI interaction with substantial human direction.
Traceability is partial, with prompts or interaction excerpts available.
The human author remains responsible for the final content.
\end{quote}

A shorter, publisher-oriented version may be:

\begin{quote}
Generative AI was used in the preparation of this text for conceptual support, drafting, and iterative revision.
The author guided the process, reviewed selected parts of the output, and remains responsible for the final content.
Partial interaction records are available.
\end{quote}

A stricter institutional version may be:

\begin{quote}
AI assistance was used at level G4 for iterative conversational generation and at level I4 for conceptual support.
The text was globally restructured at level F4 and partially reviewed by a human author at level E2.
Human control is classified as C2, and traceability is classified as T2.
\end{quote}

These three outputs serve different audiences.
The first is explanatory, the second is suitable for publisher disclosure, and the third is suitable for audit or compliance. All are generated from the same underlying annotation.

\subsection{Template-Based Generation}

Automatic prose disclosure can be implemented through a simple template system.

Each facet-level pair maps to a controlled textual fragment.

For example:

\begin{center}
\begin{tabular}{llp{6cm}}
\toprule
Code & Meaning & Textual fragment \\
\midrule
F4 & Global restructuring & The text underwent global rhetorical or structural restructuring. \\
G4 & Iterative conversation & The text was produced through iterative human-AI conversation. \\
E2 & Partial human review & The text received partial human review. \\
I4 & Conceptual support & AI was used for conceptual support, ideation, framing, or argument development. \\
C2 & Guided interaction & The process was guided by human-AI interaction with substantial human direction. \\
T2 & Partial traceability & Prompts or interaction excerpts are available. \\
\bottomrule
\end{tabular}
\end{center}

A disclosure generator can concatenate the fragments and then add a responsibility statement.

For scholarly contexts, the responsibility statement should be mandatory.

A recommended default is:

\begin{quote}
The human author remains responsible for the accuracy, integrity, originality, and final form of the text.
\end{quote}

This responsibility statement is necessary because AI identification does not transfer authorship or accountability to the system. It only clarifies the production process.

\subsection{Use in Documents}

At the document level, the annotation should appear near the title, abstract, acknowledgments, or disclosure section.

At the chapter level, it may appear immediately after the chapter title.

At the section and subsection levels, it may appear after the heading.

At the paragraph level, the iconic form may appear in the margin, while the textual grid may appear in a comment, footnote, appendix, or machine-readable metadata layer.

For example:

\begin{verbatim}
\aitextsection{F4}{G4}{E2}{I4}{C2}{T2}
\end{verbatim}

could render icons in the document and also store the textual equivalent:

\begin{center}
\texttt{|F4|G4|E2|I4|C2|T2|}
\end{center}

The same command could generate a disclosure note in an appendix.

This design allows the model to satisfy different institutional requirements without requiring authors to rewrite disclosure statements manually.

\subsection{Accessibility and Preservation}

The textual representation should always be preserved, even when icons are used. This is necessary for accessibility, indexing, search, plain-text conversion, and long-term preservation. A PDF may display icons, an HTML version may include icons with alternative text, a repository version may include the grid notation, and a publisher submission system may include the generated prose disclosure. All of these are different renderings of the same annotation.

The model, therefore, treats formal, iconic, and textual representations as complementary.While the formal representation provides precision, the iconic representation provides readability, and the textual representation provides accessibility, portability, and institutional compatibility.

\section{AI Systems as References}

AI systems should be cited as tools.

Example:

\begin{verbatim}
@misc{openai2026,
  author = {OpenAI},
  title = {GPT Models},
  year = {2026},
  url = {https://developers.openai.com},
  note = {Accessed: 2026-04-25}
}
\end{verbatim}

This supports reproducibility and transparency.

\section{Worked Example: Annotating This Article}

This section applies the proposed model to the production of the present article.

The example is included for two reasons.
First, it demonstrates how the annotation can be used in a real writing process.
Second, it shows why a binary statement such as ``AI was used'' is insufficient.

This article was produced through an extended human-AI interaction.
The human author initiated the proposal by defining the central idea: every text should be accompanied by a structured mark indicating how AI was used in its production.
The human author then specified the initial dimensions of the model, including form-related transformations, generation process, textual scale, and review status.
Across the conversation, the human author repeatedly corrected the model, rejected inadequate formulations, asked for deeper literature grounding, requested compatibility with arXiv and Overleaf, required a LaTeX style file, specified the placement of icons in the document, and refined the semantics of the facets.

The AI system contributed by drafting successive versions of the article, proposing terminology, organizing the paper structure, expanding the related work, generating examples of LaTeX commands, proposing iconographic representations, and producing explanatory text for the model.
The AI system also suggested connections with existing frameworks, including provenance models, authorship taxonomies, publisher guidelines, educational AI-use scales, and content authenticity approaches.
However, the direction of the work remained guided by the human author, who defined the research objective, requested revisions, identified conceptual problems, and selected the final interpretation of the model.

A representative document-level annotation for this article is:

\begin{equation}
S_{article} = (F4, G4, E3, I4, C2, T3)
\end{equation}

\begin{figure}
    \centering
    \includegraphics[width=6cm]{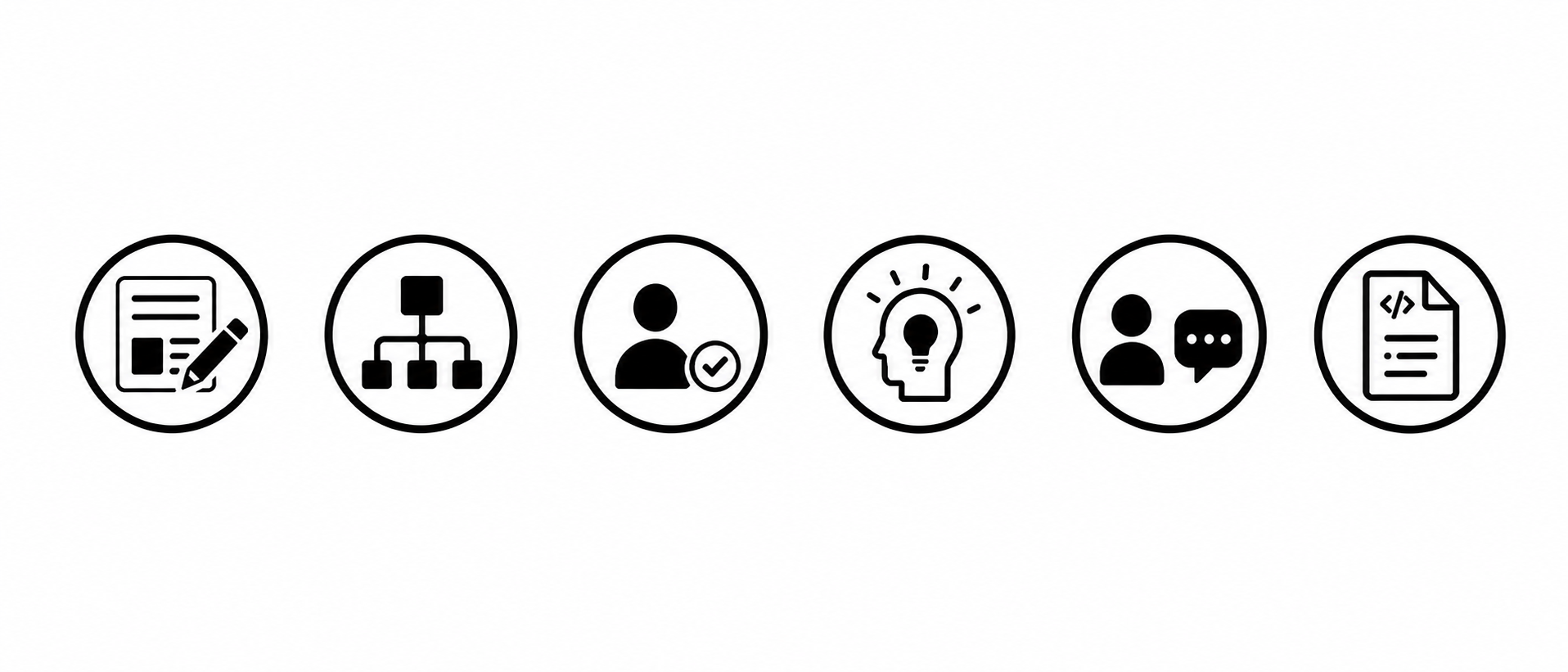}
    \caption{Classification of this article according to its proposal}
    \label{fig:example}
\end{figure}

This annotation should be read as a compact summary of the dominant production process of the article. It's iconic form is in the title page of this article and also in \autoref{fig:example}.
It does not mean that every paragraph has exactly the same classification.
Some passages were mostly generated as prose from detailed instructions.
Other passages were conceptual reorganizations requested by the human author.
Some elements, such as the icon system and LaTeX style, were developed through iterative refinement and correction.
For this reason, the annotation is best interpreted as a document-level disclosure rather than as a sentence-level provenance record.

The Form level is F4 because the article underwent global rhetorical and structural restructuring.
The text was not merely corrected for spelling or grammar.
It was repeatedly reorganized into different conceptual architectures, including a core model, an extended model, a related work section, a representation section, a worked example, and a discussion of institutional disclosure.
The revision process changed the order of sections, expanded definitions, clarified the distinction between facets, and converted brief notes into a coherent position paper.
This corresponds to global restructuring rather than local stylistic refinement.

The Generation level is G4 because the article was produced through iterative conversation.
It was not generated from a single prompt.
The human author asked for an initial proposal, then requested changes, added constraints, corrected misunderstandings, introduced new requirements, and redirected the writing several times.
For example, the Generation facet itself was revised when the human author clarified that the intended concept was conversationality rather than conservation.
Similarly, the Control facet was revised when the human author observed that its first icon was visually redundant with Generation level 0.
These episodes show that the article emerged from dialogue, correction, and progressive specification.

The Evaluation level is E3 because the article received full human review at the level of its conceptual structure and main claims.
The human author did not merely accept the AI output.
Instead, the author reviewed the proposed dimensions, identified missing sections, demanded deeper justification, corrected the organization of the model, required a more careful related work review, and requested changes to visual representations.
This is stronger than partial review because the main architecture of the article was repeatedly inspected and redirected.
At the same time, it is not classified as E4 because the article has not yet undergone independent external validation, formal peer review, or multi-reviewer assessment.

The Intent level is I4 because AI was used for conceptual support, ideation, framing, and argument development.
The AI system was not used only for grammar correction or stylistic polishing.
It contributed to the articulation of the model, the grouping of related work, the formulation of distinctions between core and extended facets, and the construction of explanatory examples.
Although the human author retained control over the research direction, the AI system participated in shaping the conceptual language of the paper.
This places the article at the highest Intent level in the proposed scale.

The Control level is C2 because the process was a guided human-AI interaction.
The human author set the goals, imposed constraints, evaluated outputs, and required corrections.
The AI system generated substantial text and proposed structures, but it did not autonomously define the project.
The process was therefore neither fully human-controlled in the narrow sense of C0 nor AI-dominant in the sense of C3.
It is best characterized as guided interaction, where the human author directed the project and the AI system acted as a generative and organizational collaborator.

The Traceability level is T3 because the conversation provides a substantial record of prompts, intermediate outputs, corrections, model-driven revisions, and design decisions.
The production process can be reconstructed more fully than in a case where only a final disclosure statement is available.
The conversation includes evidence of changes in the model, such as the transition from the initial three facets to the extended model, the introduction of textual and iconic representations, the request for publisher-oriented disclosure, and the worked example itself.
However, the annotation is not T4 because the record is not independently verified through signed metadata, authenticated logs, cryptographic provenance, repository history, or a third-party archival mechanism.

The same annotation can be expressed in textual grid form as:

\begin{center}
\texttt{|F4|G4|E3|I4|C2|T3|}
\end{center}

A publisher-oriented disclosure generated from this annotation could be:

\begin{quote}
This article was prepared with substantial assistance from a generative AI system.
AI was used through iterative human-AI conversations for conceptual support, drafting, restructuring, and explanatory refinement.
The human author guided the process, reviewed the complete conceptual structure and main claims, corrected inadequate outputs, and remains responsible for the final text.
The production process is substantially traceable through prompts, intermediate responses, and revision history, although it has not been independently verified through external provenance mechanisms.
\end{quote}

A shorter version suitable for a submission form could be:

\begin{quote}
Generative AI was used for conceptual support, drafting, and iterative revision.
The author guided the process, fully reviewed the manuscript's structure and main claims, and remains responsible for the final content.
Prompts and intermediate interaction records are available.
\end{quote}

This example illustrates the purpose of the model.
The article cannot be accurately described as simply human-written or AI-written.
It is a hybrid artifact produced through conceptual direction by a human author, extensive AI-assisted drafting, iterative conversational refinement, human review, and preserved interaction records.
The faceted annotation makes this hybrid process explicit without reducing it to a single misleading category.

\section{Conclusion}

This paper proposes a faceted model for the transparent attribution of AI-assisted text production.

The starting point of the proposal is the observation that the current language used to declare generative AI assistance in academic and scientific writing is still too imprecise for the problems it is expected to solve.
A statement such as ``AI was used in manuscript preparation'' may be acceptable as a minimal act of disclosure, but it does not tell the reader what kind of intervention occurred, where it occurred, how extensive it was, or whether the resulting text was reviewed.
It also does not distinguish between a passage corrected for spelling, a paragraph rewritten for style, a section generated from a prompt, an argument developed through iterative conversation, a bibliography formatted by an assistant, or a conceptual model partially shaped through dialog with a language model.
For this reason, the paper argues that the central problem is not simply whether AI should be disclosed, because many institutions already require it.
The deeper problem is how AI use should be described once disclosure becomes necessary.

The main contribution of the paper is therefore a vocabulary for structured disclosure.
The proposal separates AI-assisted text production into facets, because no single scale can capture the relevant distinctions.
The core model defines Form, Generation, and Evaluation as a minimal operational structure.
Form describes what happened to the linguistic and rhetorical shape of the text after an initial version existed.
Generation describes how the text came into being, including whether it was human-authored, completed by AI, generated from a prompt, developed through conversation, or produced through a structured pipeline.
Evaluation describes whether the result was accepted without review, checked by automated systems, partially reviewed by a human, fully reviewed by a human, or validated through a multi-stage process.
These three facets already make it possible to distinguish cases that are often collapsed into a single misleading category. 

The extended model adds Intent, Control, and Traceability because minimal disclosure is useful, but it is not always sufficient.
Intent describes why AI was used, which matters because linguistic correction, translation, summarization, drafting, and conceptual support have different implications for authorship and responsibility.
Control describes who directed the process, which matters because a human-guided dialog is different from AI-dominant production with only light human approval.
Traceability describes whether the process can be reconstructed, which matters because a claim about AI use is stronger when prompts, logs, model identifiers, intermediate versions, or provenance records are available.
Together, these extended facets make the model more expressive without forcing every user to adopt maximum detail in every context.
This distinction between a core model and an extended model is important.
If the model is too simple, it will reproduce the vagueness of current disclosure statements.
If it is too detailed, it will become burdensome, inconsistent, and unlikely to be adopted.
The proposed structure, therefore, tries to balance usability and fidelity.

The paper also contributes a representational distinction that is essential for practical adoption.
The annotation itself is semantic, but it can be rendered in different ways.
A formal tuple provides precision.
An iconic representation provides rapid visual recognition.
A textual grid provides portability, accessibility, and compatibility with plain-text environments.
A generated prose statement provides a form suitable for publishers, universities, funding agencies, journals, and submission systems.
This separation between meaning and rendering allows the same annotation to appear in different institutional and technical contexts.
For example, a PDF may display icons, an HTML version may include alternative text, a repository may store the grid notation, and a publisher form may receive an automatically generated disclosure paragraph.
The underlying annotation can remain stable while its expression changes according to the medium.

This representational strategy has historical precedents in earlier Web metadata systems.
The Platform for Internet Content Selection, or PICS, showed that Web resources could be associated with structured labels defined by rating systems and distributed in different ways, including embedding in HTML, transmission through protocol headers, or distribution separately from the document \citep{w3c1996picslabels}.
The purpose of PICS was not scholarly AI disclosure, but its architecture remains instructive.
It separated resources from the labels that described them and allowed different agents to interpret those labels according to local rules.
Later metadata systems, including RDF and POWDER, generalized the idea that resources and groups of resources can be described through structured, machine-readable descriptions \citep{w3c2014rdf, w3c2009powderdr, w3c2009powderformal}.
The present proposal can be understood as a step in a similar direction.
It does not claim to define a complete Web-scale infrastructure for AI-use metadata, but it proposes the semantic categories that such an infrastructure would need if it were to describe AI-assisted scholarly writing with more precision than ordinary prose disclosure.

The worked example in this paper illustrates why such precision matters.
The article itself is not accurately described as simply human-written or AI-written.
It was produced through human conceptual direction, AI-assisted drafting, iterative conversational refinement, human review, and preserved interaction records.
The document-level annotation \texttt{|F4|G4|E3|I4|C2|T3|} captures this hybrid production process more clearly than a conventional disclosure statement.
It indicates that the article underwent global restructuring, was produced through iterative conversation, received full human review at the level of conceptual structure and main claims, used AI for conceptual support, remained under guided human control, and preserved substantial traceability through the conversation record.
Even so, the annotation remains a simplification.
Different sections of the same article may have different production histories.
A formal definition, a related work paragraph, an icon caption, a bibliographic note, and a worked example may each involve different degrees and kinds of AI assistance.
This is why the model is designed for multiple textual scales rather than only for whole-document disclosure.

The broader importance of the model comes from the status of generative AI as a durable cognitive tool.
AI-assisted writing is unlikely to disappear from academic and scientific communication.
The relevant question is therefore not whether scholars will use AI, but how scholarly communities will make that use visible, accountable, interpretable, and reviewable.
Earlier cognitive tools changed scholarship by extending perception, calculation, memory, search, simulation, communication, and symbolic manipulation.
Generative AI now extends linguistic production, conceptual exploration, reformulation, summarization, translation, and argumentative drafting.
However, this extension reaches activities that have traditionally been treated as signs of authorship, originality, reasoning, and intellectual responsibility.
This is why vague disclosure is inadequate.
It is also why simple prohibition is unlikely to be a sufficient long-term response.
A mature scholarly response requires instruments that make AI-assisted cognition describable without reducing the text to a binary opposition between human and machine.

For these reasons, this paper should be read as a call to action rather than as a finished standard.
The proposed facets and levels need empirical testing.
Researchers should examine whether different annotators can apply the model consistently to the same writing process.
Authors and editors should test whether the model is usable in real workflows.
Tool builders should explore implementations in LaTeX, word processors, repository systems, browser extensions, and publisher submission platforms.
Standards-oriented researchers should examine how the model could be represented in PROV, RDF, C2PA-style metadata, or other provenance infrastructures.
Educators should test whether the model helps students distinguish assistance, delegation, revision, authorship, and responsibility.
Policy makers should determine which facets should be required, recommended, or optional in different contexts.

The model should also remain open to refinement.
Future versions may need additional facets, such as source grounding, domain risk, rights status, audience, privacy sensitivity, data sensitivity, model identity, or evidentiary basis.
However, expansion should be cautious.
Each new facet increases expressive power, but it also increases the cost of annotation.
A successful community standard should not attempt to encode every possible distinction from the beginning.
It should preserve the distinctions that matter most for interpretation, responsibility, and reproducibility, while leaving room for extensions in high-stakes or specialized contexts.
In this sense, the distinction between the core model and the extended model should be preserved.
The core model should remain small enough to encourage adoption.
The extended model should remain available when stronger accountability is required.

The immediate task is practical as much as theoretical.
Scholarly communities need to move from informal confession to structured disclosure.
They need ways to say not only that AI was used, but how it was used, why it was used, who controlled the process, how the output was evaluated, and whether the process can be reconstructed.
The model proposed here offers one possible starting point.
Its value will depend on whether it can be criticized, simplified, extended, implemented, and tested in real publication and educational contexts.
If it succeeds, it will not be because the present paper defines a final taxonomy.
It will be because the paper helps establish a shared descriptive language for a problem that is becoming central to the future of academic and scientific communication.

\printbibliography

@misc{c2pa2024,
  author = {C2PA},
  title = {C2PA Specification},
  year = {2024},
  url = {https://c2pa.org}
}

@article{thorp2023chatgpt,
  author = {Thorp, H. Holden},
  title = {ChatGPT is fun, but not an author},
  journaltitle = {Science},
  volume = {379},
  number = {6630},
  pages = {313},
  date = {2023},
  doi = {10.1126/science.adg7879}
}

@article{nature2023transparent,
  author = {{Nature}},
  title = {Tools such as ChatGPT threaten transparent science: Here are our ground rules for their use},
  journaltitle = {Nature},
  volume = {613},
  number = {7945},
  pages = {612},
  date = {2023},
  doi = {10.1038/d41586-023-00191-1}
}

@article{stokelwalker2023chatgpt,
  author = {Stokel-Walker, Chris and Van Noorden, Richard},
  title = {What ChatGPT and generative AI mean for science},
  journaltitle = {Nature},
  volume = {614},
  number = {7947},
  pages = {214--216},
  date = {2023},
  doi = {10.1038/d41586-023-00340-6}
}

@article{walters2023fabrication,
  author = {Walters, William H. and Wilder, Esther I.},
  title = {Fabrication and errors in the bibliographic citations generated by ChatGPT},
  journaltitle = {Scientific Reports},
  volume = {13},
  number = {1},
  pages = {14045},
  date = {2023},
  doi = {10.1038/s41598-023-41032-5}
}

@article{emsley2023fabrications,
  author = {Emsley, Robin},
  title = {ChatGPT: These are not hallucinations: They are fabrications and falsifications},
  journaltitle = {Schizophrenia},
  volume = {9},
  number = {1},
  pages = {52},
  date = {2023},
  doi = {10.1038/s41537-023-00379-4}
}

@article{sun2024hallucination,
  author = {Sun, Yifei and Sheng, Di and Zhou, Zhiheng and Wu, Yujia},
  title = {AI hallucination: Towards a comprehensive classification of distorted information in artificial intelligence-generated content},
  journaltitle = {Humanities and Social Sciences Communications},
  volume = {11},
  pages = {1278},
  date = {2024},
  doi = {10.1057/s41599-024-03811-x}
}

@article{bender2021dangers,
  author = {Bender, Emily M. and Gebru, Timnit and McMillan-Major, Angelina and Shmitchell, Shmargaret},
  title = {On the dangers of stochastic parrots: Can language models be too big?},
  journaltitle = {Proceedings of the 2021 ACM Conference on Fairness, Accountability, and Transparency},
  pages = {610--623},
  date = {2021},
  doi = {10.1145/3442188.3445922}
}

@online{weidinger2021ethical,
  author = {Weidinger, Laura and Mellor, John and Rauh, Maribeth and Griffin, Conor and Uesato, Jonathan and Huang, Po-Sen and Cheng, Myra and Glaese, Amelia and Balle, Borja and Kasirzadeh, Atoosa and Kenton, Zachary and Brown, Sasha and Hawkins, Will and Stepleton, Tom and Biles, Courtney and Birhane, Abeba and Haas, Julia and Rimell, Laura and Hendricks, Lisa Anne and Isaac, William and Legassick, Sophie and Irving, Geoffrey and Gabriel, Iason},
  title = {Ethical and social risks of harm from language models},
  date = {2021},
  eprint = {2112.04359},
  eprinttype = {arxiv},
  eprintclass = {cs.CL}
}

@online{wired2023academicai,
  author = {Knibbs, Kate},
  title = {Use of AI is seeping into academic journals, and it is proving difficult to detect},
  organization = {Wired},
  date = {2023-08-15},
  url = {https://www.wired.com/story/use-of-ai-is-seeping-into-academic-journals-and-its-proving-difficult-to-detect/},
  urldate = {2026-04-25}
}

@online{times2025fakecitations,
  author = {Woolcock, Nicola},
  title = {Publisher under fire after fake citations found in AI ethics guide},
  organization = {The Times},
  date = {2025-12-14},
  url = {https://www.thetimes.com/uk/science/article/ai-ethics-guide-citations-nsnjmz25b},
  urldate = {2026-04-25}
}

@online{elsevier2025genai,
  author = {{Elsevier}},
  title = {Generative AI policies for journals},
  date = {2025-09},
  url = {https://www.elsevier.com/about/policies-and-standards/generative-ai-policies-for-journals},
  urldate = {2026-04-25}
}

@online{springernature2026ai,
  author = {{Springer Nature}},
  title = {Artificial Intelligence: Editorial policies},
  date = {2026},
  url = {https://www.nature.com/nature-portfolio/editorial-policies/ai},
  urldate = {2026-04-25}
}

@online{acm2025authorship,
  author = {{Association for Computing Machinery}},
  title = {ACM Policy on Authorship},
  date = {2025-09-16},
  url = {https://www.acm.org/publications/policies/new-acm-policy-on-authorship},
  urldate = {2026-04-25}
}

@online{harvard2026genai,
  author = {{Harvard University}},
  title = {Research with Generative AI},
  date = {2026},
  url = {https://www.harvard.edu/ai/research-resources/},
  urldate = {2026-04-25}
}

@online{ufrj2026ai,
  author = {{Universidade Federal do Rio de Janeiro}},
  title = {UFRJ atualiza diretrizes de integridade acadêmica e propõe recomendações sobre o uso de inteligência artificial},
  date = {2026-03-18},
  url = {https://conexao.ufrj.br/2026/03/ufrj-atualiza-diretrizes-de-integridade-academica-e-propoe-recomendacoes-sobre-o-uso-de-inteligencia-artificial/},
  urldate = {2026-04-25}
}

@online{cnpq2026integridade,
  author = {{Conselho Nacional de Desenvolvimento Científico e Tecnológico}},
  title = {CNPq institui Política de Integridade na Atividade Científica, que estabelece normas e boas práticas de atuação},
  date = {2026-03-11},
  url = {https://www.gov.br/cnpq/pt-br/assuntos/noticias/cnpq-em-acao/cnpq-publica-portaria-que-institui-politica-de-integridade-na-atividade-cientifica},
  urldate = {2026-04-25}
}

@report{usco2025copyrightability,
  author = {{United States Copyright Office}},
  title = {Copyright and Artificial Intelligence, Part 2: Copyrightability},
  institution = {United States Copyright Office},
  date = {2025-01},
  url = {https://www.copyright.gov/ai/Copyright-and-Artificial-Intelligence-Part-2-Copyrightability-Report.pdf},
  urldate = {2026-04-25}
}

@online{wipo2024genaiip,
  author = {{World Intellectual Property Organization}},
  title = {Generative AI: Navigating intellectual property},
  date = {2024},
  url = {https://www.wipo.int/publications/en/details.jsp?id=4713},
  urldate = {2026-04-25}
}

@misc{moreau2013prov,
  author = {Moreau, Luc and Missier, Paolo},
  title = {{PROV-DM}: The {PROV} Data Model},
  date = {2013},
  organization = {World Wide Web Consortium},
  url = {https://www.w3.org/TR/prov-dm/},
  urldate = {2026-04-25}
}

@article{allen2014credit,
  author = {Allen, Liz and Scott, Jo and Brand, Amy and Hlava, Marjorie and Altman, Micah},
  title = {Publishing: Credit where credit is due},
  journaltitle = {Nature},
  volume = {508},
  number = {7496},
  pages = {312--313},
  date = {2014},
  doi = {10.1038/508312a}
}

@article{brand2015beyond,
  author = {Brand, Amy and Allen, Liz and Altman, Micah and Hlava, Marjorie and Scott, Jo},
  title = {Beyond authorship: Attribution, contribution, collaboration, and credit},
  journaltitle = {Learned Publishing},
  volume = {28},
  number = {2},
  pages = {151--155},
  date = {2015},
  doi = {10.1087/20150211}
}

@inproceedings{mitchell2019model,
  author = {Mitchell, Margaret and Wu, Simone and Zaldivar, Andrew and Barnes, Parker and Vasserman, Lucy and Hutchinson, Ben and Spitzer, Elena and Raji, Inioluwa Deborah and Gebru, Timnit},
  title = {Model Cards for Model Reporting},
  booktitle = {Proceedings of the Conference on Fairness, Accountability, and Transparency},
  pages = {220--229},
  date = {2019},
  doi = {10.1145/3287560.3287596}
}

@article{gebru2021datasheets,
  author = {Gebru, Timnit and Morgenstern, Jamie and Vecchione, Briana and Vaughan, Jennifer Wortman and Wallach, Hanna and Daum{\'e} III, Hal and Crawford, Kate},
  title = {Datasheets for Datasets},
  journaltitle = {Communications of the ACM},
  volume = {64},
  number = {12},
  pages = {86--92},
  date = {2021},
  doi = {10.1145/3458723}
}

@online{cope2023aitools,
  author = {{Committee on Publication Ethics}},
  title = {Authorship and AI Tools},
  date = {2023-02-13},
  url = {https://publicationethics.org/guidance/cope-position/authorship-and-ai-tools},
  urldate = {2026-04-25}
}

@online{plos2023ai,
  author = {{PLOS}},
  title = {Research Integrity and Publication Ethics: Use of AI Writing Tools},
  date = {2023},
  url = {https://plos.org/research-integrity-and-ethics/},
  urldate = {2026-04-25}
}

@online{ieeeras2026genai,
  author = {{IEEE Robotics and Automation Society}},
  title = {Guidelines for Generative AI Usage},
  date = {2026},
  url = {https://www.ieee-ras.org/publications/guidelines-for-generative-ai-usage/},
  urldate = {2026-04-25}
}

@article{weaver2024aid,
  author = {Weaver, Kari D.},
  title = {The Artificial Intelligence Disclosure ({AID}) Framework: An Introduction},
  journaltitle = {College \& Research Libraries News},
  volume = {85},
  number = {8},
  date = {2024},
  url = {https://crln.acrl.org/index.php/crlnews/article/view/26548},
  urldate = {2026-04-25}
}

@article{weaver2026aid,
  author = {Weaver, Kari D.},
  title = {Artificial Intelligence and Transparency: Toward a Framework for Disclosure of AI Use in Learning, Research, and Publication},
  journaltitle = {Information Services \& Use},
  date = {2026},
  doi = {10.1177/18758789261435716}
}

@online{gaidet2025,
  author = {Teixeira da Silva, Jaime A. and Suchikova, Yana},
  title = {{GAIDeT}: A Practical Taxonomy for Declaring AI Use in Research and Publishing},
  date = {2025-08-25},
  organization = {Leiden Madtrics},
  url = {https://www.leidenmadtrics.nl/articles/gaidet-a-practical-taxonomy-for-declaring-ai-use-in-research-and-publishing},
  urldate = {2026-04-25}
}

@online{stm2023genai,
  author = {{International Association of Scientific, Technical and Medical Publishers}},
  title = {Generative AI in Scholarly Communications: Ethical and Practical Guidelines for the Use of Generative AI in the Publication Process},
  date = {2023-12-19},
  url = {https://stm-assoc.org/document/stm-generative-ai-paper-2023/},
  urldate = {2026-04-25}
}

@online{stm2025classification,
  author = {{International Association of Scientific, Technical and Medical Publishers}},
  title = {New STM Draft Report: Classifying AI Use in Manuscript Preparation},
  date = {2025-04-21},
  url = {https://stm-assoc.org/new-stm-draft-report-classifying-ai-use-in-manuscript-preparation/},
  urldate = {2026-04-25}
}

@article{perkins2024aias,
  author = {Perkins, Mike and Furze, Leon and Roe, Jasper and MacVaugh, Jason},
  title = {The Artificial Intelligence Assessment Scale ({AIAS}): A Framework for Ethical Integration of Generative AI in Educational Assessment},
  journaltitle = {Journal of University Teaching and Learning Practice},
  volume = {21},
  number = {6},
  date = {2024},
  doi = {10.53761/q3azde36}
}

@article{resnik2025disclosing,
  author = {Resnik, David B. and Hosseini, Mohammad},
  title = {Disclosing Artificial Intelligence Use in Scientific Research and Publication},
  journaltitle = {Accountability in Research},
  date = {2025},
  doi = {10.1080/08989621.2025.2481949}
}

@misc{daisy2026,
  title = {AI Disclosure with {DAISY}},
  date = {2026},
  eprint = {2604.02760},
  eprinttype = {arxiv},
  eprintclass = {cs.HC},
  url = {https://arxiv.org/abs/2604.02760},
  urldate = {2026-04-25}
}

@misc{liang2024monitoring,
  author = {Liang, Weixin and Izzo, Zachary and Zhang, Yuhui and Lepp, Hannah and Cao, Hancheng and Zhao, Xuandong and Chen, Lingjiao and Ye, Haotian and Liu, Sheng and Huang, Zhi and others},
  title = {Monitoring AI-Modified Content at Scale: A Case Study on the Impact of ChatGPT on AI Conference Peer Reviews},
  date = {2024},
  eprint = {2403.07183},
  eprinttype = {arxiv},
  eprintclass = {cs.CL}
}

@online{knibbs2023academicai,
  author = {Knibbs, Kate},
  title = {Use of AI Is Seeping Into Academic Journals, and It Is Proving Difficult to Detect},
  organization = {Wired},
  date = {2023-08-15},
  url = {https://www.wired.com/story/use-of-ai-is-seeping-into-academic-journals-and-its-proving-difficult-to-detect/},
  urldate = {2026-04-25}
}

@article{perlis2026disclosure,
  author = {Perlis, Roy H. and others},
  title = {Author Disclosure of Use of AI in Submissions to 13 {JAMA} Network Journals},
  journaltitle = {JAMA},
  date = {2026},
  doi = {10.1001/jama.2025.25300}
}

@misc{w3c1996picslabels,
  author = {{World Wide Web Consortium}},
  title = {{PICS} Label Distribution Label Syntax and Communication Protocols},
  year = {1996},
  month = oct,
  howpublished = {\url{https://www.w3.org/TR/REC-PICS-labels-961031}},
  note = {Accessed: 2026-04-25}
}

@misc{w3c2014rdf,
  author = {{World Wide Web Consortium}},
  title = {{RDF} 1.1 Concepts and Abstract Syntax},
  year = {2014},
  month = feb,
  howpublished = {\url{https://www.w3.org/TR/rdf11-concepts/}},
  note = {Accessed: 2026-04-25}
}

@misc{w3c2009powderdr,
  author = {Archer, Phil and Smith, Kevin and Perego, Andrea},
  title = {Protocol for Web Description Resources ({POWDER}): Description Resources},
  year = {2009},
  month = sep,
  howpublished = {\url{https://www.w3.org/TR/powder-dr/}},
  note = {W3C Recommendation. Accessed: 2026-04-25}
}

@misc{w3c2009powderformal,
  author = {Archer, Phil and Smith, Kevin and Perego, Andrea},
  title = {Protocol for Web Description Resources ({POWDER}): Formal Semantics},
  year = {2009},
  month = sep,
  howpublished = {\url{https://www.w3.org/TR/powder-formal/}},
  note = {W3C Recommendation. Accessed: 2026-04-25}
}

\end{document}